# Disposable face masks: a direct source for inhalation of microplastics

Andres F. Prada[1*], Avram Distler[2], Shyuan Cheng[2], John W. Scott[1], Leonardo P. Chamorro[2,3,4,5], Ganesh Subramanian[4], Vishal Verma[4], Andrew Turner[6]

[1] *Illinois Sustainable Technology Center, Prairie Research Institute, University of Illinois at Urbana-Champaign, Champaign, IL 61820, USA*
[2] *Mechanical Science and Engineering, University of Illinois at Urbana-Champaign, Urbana, IL 61801, USA*
[3] *Aerospace Engineering, University of Illinois at Urbana-Champaign, Urbana, IL 61801, USA*
[4] *Civil and Environmental Engineering, University of Illinois at Urbana-Champaign, Urbana, IL 61801, USA*
[5] *Geology, University of Illinois at Urbana-Champaign, Urbana, IL 61801, USA*
[6] *School of Geography, Earth and Environmental Sciences, University of Plymouth, Drake Circus, Plymouth PL4 8AA, UK*
[*] Corresponding author. Email: pradase2@illinois.edu

**Abstract**
Surgical masks have played a crucial role in healthcare facilities to protect against respiratory and infectious diseases, particularly during the COVID-19 pandemic. However, the synthetic fibers, mainly made of polypropylene, used in their production may adversely affect the environment and human health. Recent studies have confirmed the presence of microplastics and fibers in human lungs and have related these synthetic particles with the occurrence of pulmonary ground glass nodules. Using a piston system to simulate human breathing, this study investigates the role of surgical masks as a direct source of inhalation of microplastics. Results reveal the release of particles of sizes ranging from nanometers (300 nm) to millimeters (~2 mm) during normal breathing conditions, raising concerns about the potential health risks. Notably, large visible particles (> 1 mm) were observed to be ejected from masks with limited wear after only a few breathing cycles. Given the widespread use of masks by healthcare workers and the potential future need for mask usage by the general population during seasonal infectious diseases or new pandemics, developing face masks using safe materials for both users and the environment is imperative.

**Keywords:** Synthetic fibers, polypropylene, plastic pollution, human exposure, laboratory experiments.

## 1. Introduction

During and since the COVID-19 pandemic, the use of face masks by the general population has increased exponentially to reduce the spread of the highly contagious SARS-CoV-2 virus. Beginning in mid-2020, masks became mandatory for over a year in many countries in Europe and America for indoor use in situations such as shopping, classrooms, or public transportation, although in countries like China, people have been wearing masks in public for decades (Matuschek et al., 2020). As it was not common in Europe and America to wear a mask daily, the supply of face masks was limited, and many nations experienced shortages of these products (Bayersdorfer et al., 2020). Modern face masks are disposable and designed with specific filtration qualities for health care situations (i.e., surgical masks). Surgical masks are intended for single use to avoid sterilization, reduce labor costs, and facilitate management supplies in healthcare facilities (Strasser and Schlich, 2020).

Disposable face masks were part of the hospital care transformation in the 1960s when the system transitioned towards total disposable supplies, including syringes, needles, trays, and surgical instruments (Strasser and Schlich, 2020). In the previous decades, the international surgery community had used washable and sterilizable masks made with layers of cotton gauze (Matuschek et al., 2020). Since the 1960s, however, masks have been manufactured with non-woven fabrics made of synthetic fibers that can be used only once because they deteriorate during sterilization (Strasser and Schlich, 2020) and may have an impact on human health.

The synthetic fibers used to manufacture surgical masks are commonly made of polypropylene (Tcharkhtchi et al., 2021). Polypropylene is a synthetic polymer (plastic) widely used for several applications and consumer products due to its low cost, processability, and mechanical integrity. It has hydrophobic (nonabsorbent) properties (Akalin et al., 2010), and in the case of a surgical masks, it can be treated with dimethyl-dioctadecyl-ammonium bromide to impart a positive electrical charge that attracts airborne particles (Kang and Shah, 1997). Other synthetic polymers can also be found in masks, such as polyethylene, polyesters, polyamides, polycarbonates, and polyphenylene oxide (Tcharkhtchi et al., 2021). Commonly, masks are designed with a three-layer structure: (1) the outer layer, which repels exterior humidity; (2) the middle layer, which filters airborne particles; and (3) the inner layer, which absorbs the breathing moisture (Yao et al., 2019). However, N95 respirators have one additional middle layer to provide extra support (Ju et al., 2021).

Each layer in a mask is a non-woven synthetic fabric of different thicknesses, fiber diameters, and pore sizes. Manufacturing non-woven fabrics is easy and inexpensive, making it convenient for mass production (Ju et al., 2021). Each layer has a thickness of tens of micrometers (μm) (Shokri et al., 2015), allowing the air to pass through the fabric instead of through the mask edges. The outer layer has fibers of a diameter of 20 μm and pore sizes of up to 100 μm preventing aqueous droplets from entering and damaging the filter (Zhao et al., 2020). The filter (i.e., the middle layer) has fibers of diameter 1–10 µm and pore sizes of up to 20 µm (Zhao et al., 2020) and is capable of capturing particles > 0.3 µm by impact or sedimentation and particles < 0.2 µm by diffusion and electrostatic attraction (O'Dowd et al., 2020).

Food and Drug Administration (FDA) regulates surgical masks in the United States, and they must undergo standard tests to evaluate their effectiveness in particle and bacterial/viral filtration, fluid resistance, and breathability (Ju et al., 2021). The particle filtration efficiency of an N95 respirator is generally > 95%, while a surgical mask is usually between 19 to 33% (Zhao et al., 2020). One condition that is not tested is the number of fibers that can be released from the mask during breathing, which constitutes direct exposure to the inhalation of synthetic fibers (Li et al., 2021). Besides the ubiquity of microfibers in outdoor and indoor spaces, which have been reported as being found suspended in the air and may enter the human body through the respiratory tract (Vianello et al., 2019; Zhang et al., 2020), the use of surgical masks increases considerably the chances of microplastic intake during breathing.

Li et al. (2021) first warned of the microplastic inhalation risk of wearing masks. They tested different new and reused masks under a suction pump that generated a constant unidirectional flow for prolonged times from 2 to 720 hours. In this study, we developed an oscillatory airflow system to recreate human breathing conditions and showed, with the help of a high-speed camera, that fibers can be released from a new surgical mask in a few seconds under the oscillatory flow. We further recorded and counted smaller particles (0.3 – 10µm) using an optical particle sizer attached to the airflow system.

2. **Materials and Methods**

A customized piston was built to generate an oscillatory airflow based on the design used by Yuk et al. (2022). The piston (Figure 1a) comprises a transparent PVC pipe with a diameter of 100 mm and a length of 500 mm (part number 49035K51, McMaster-Carr Supply Co.). A 75-mm PVC pipe is housed within this outer pipe, featuring an end plate that functions as the moving piston. The outer pipe is then connected to a 25-mm transparent PVC outlet pipe, enhancing the speed of the exiting airflow. A straight reducer with a 100-mm to 25-mm diameter (part number 4880K017, McMaster-Carr Supply Co.) facilitates the connection between the two pipes. The 25-mm outlet pipe extends by 80 mm, and a 50 mm square duct is welded to hold the mask securely and improve visibility during camera recordings (Figure 1b). The piston has a stroke length of 360 mm and is affixed to frictionless bearings linked to acrylic components measuring 320 mm and 180 mm in length. The motion of the piston is driven by a stepper motor (part number 34MDSI314S, Anaheim Automation), which is controlled by software designed for motor step speed regulation.

A series of initial experiments were performed using a new mask. The mask was exposed to an oscillatory airflow system operating at a frequency of 0.25 Hz, resulting in a volumetric flow rate of Q = 45 L/min. The maximum airflow speed reached approximately 1.6 m/s in both directions at the exit of the 25-mm outlet pipe. The volumetric flow rate of 45 L/min falls within the typical range of average human breathing in adults (approximately 15 to 85 L/min, as reported in studies by He et al. (2013) and Basu (2021)).

To capture the behavior of the mask under the airflow system, a high-speed camera model Phantom Miro M340, equipped with a 60 mm Nikon lens, was positioned at a distance of 250 mm from the square duct that held the mask. Videos of the mask's response to the airflow were obtained at a rate of 50 frames per second (fps), in a selected field of view (FOV) measuring 56 mm × 31 mm.

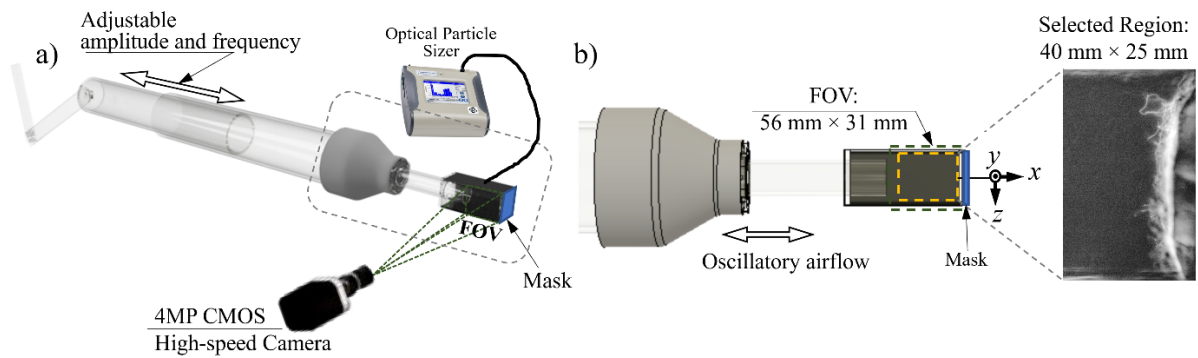

**Figure 1**. Schematics of the oscillatory airflow system and experimental setup. a) The piston with the location of the high-speed camera for fiber shedding visualization. b) Field of view (FOV) of the camera and selected region of interest.

The new surgical masks were procured from an FDA-approved supplier in the United States. The supplier boasts the world's largest face mask manufacturing facility, with a daily production capacity of 10 million units. These masks are composed of three layers of polypropylene fibers. The polymer type was confirmed using infrared spectroscopy obtaining a 79% match with reference material from the Hummel polymer sample library.

An Optical Particle Sizer (OPS 3330, TSI Incorporated) was attached to the square duct for particle analysis. This instrument enabled the measurement of particle number concentration (PNC) and particle size distribution at a frequency of 1 Hz. The OPS can detect particles in size range of 0.3 to 10 μm, providing real-time size distribution and PNC data in particles/cm³. Furthermore, it has a broad detection range from 0 to 3,000 particles/cm³.

## 3. Results and discussion

The ejection of fibers from the face masks is visually demonstrated through high-speed images captured by the camera and supported by the information on PNC and size distribution collected with the OPS. Figure 2 depicts a specific event captured once out of many trials when a fiber is released from a new mask after 3.3 seconds from initiating the oscillatory airflow system. Figure 3 summarizes the concentrations and particle size distributions measured by the OPS.

During the operation of the airflow system without a mask, we measured the average ambient PNC in the laboratory room, denoted as $C_0$, which was 19.6 particles/cm³. We used this baseline concentration ($C_0$) to normalize other PNC measurements during the actual experiment, $C$. Interestingly, when comparing the PNC with the system on and off in the room without a mask, we observed no significant difference ($C/C_0 \sim 1$), as depicted in Figure 3a, although some fluctuations occurred probably due to the changes in airflow direction after turning on the system. Furthermore, the size distributions of particles remained consistent between these two scenarios, as shown in Figure 3b. These results indicate that the mechanical system did not generate additional particles during operation.

Upon wearing the mask, we noted a reduction in the PNC of ambient particles by approximately 50% ($C/C_0 \sim 0.5$), as illustrated in Figure 3c, when the system was off. This suggests that the filtration efficiency of the mask to capture ambient particles is about 50%. However, once the system was turned on, we observed a subsequent increase in the PNC of around 10% ($C/C_0 \sim 0.6$), indicating that the mask started generating particles into the system. This increase could be due to fiber shedding or resuspension of filtered particles from the masks, induced by the mechanical airflow. This finding is further supported by the change in particle size distribution, with a noticeable increase in the percentage of particles with sizes of 0.3 μm when the mask was worn under the specified flow rate (Figure 3d). The increase in the PNC during breathing action is predominantly attributed to an increase in the PNC of particles under 0.65 μm, which was accompanied by a slight reduction in the PNC of relatively larger particles over 0.65 μm. This decrease (for particles > 0.65 μm) could be explained by the enhanced impaction of larger-sized particles onto the mask fibers due to higher airflow velocity during the breathing action.

The continuous release of synthetic particles from these masks raises concerns as these particles, particularly those smaller than 2.5 μm can be inhaled, and travel through the respiratory tract, ultimately reaching the lungs. Several studies (Pauly et al., 1998; Huang et al., 2022; Jenner et al., 2022) have confirmed the presence of microplastics in human lungs with sizes < 500 μm. However, the exact effects of microplastic inhalation on human health remain largely unknown. While the chronic effects of inhaling mineral fibers, such as asbestos (Spasiano and Pirozzi, 2017; Yang et al., 2018), in heavily exposed industrial workers or inhaling particulate matter like PM2.5 from environmental pollution have been extensively studied (Donaldson et al., 2002), little is known about the consequences of the deposition of non-degradable synthetic polymers in the lungs. Some studies have found a statistical correlation between the presence of microplastics in the lungs, impaired pulmonary function, and the development of lung ground glass nodules (Atis et al., 2005; Chen et al., 2022).

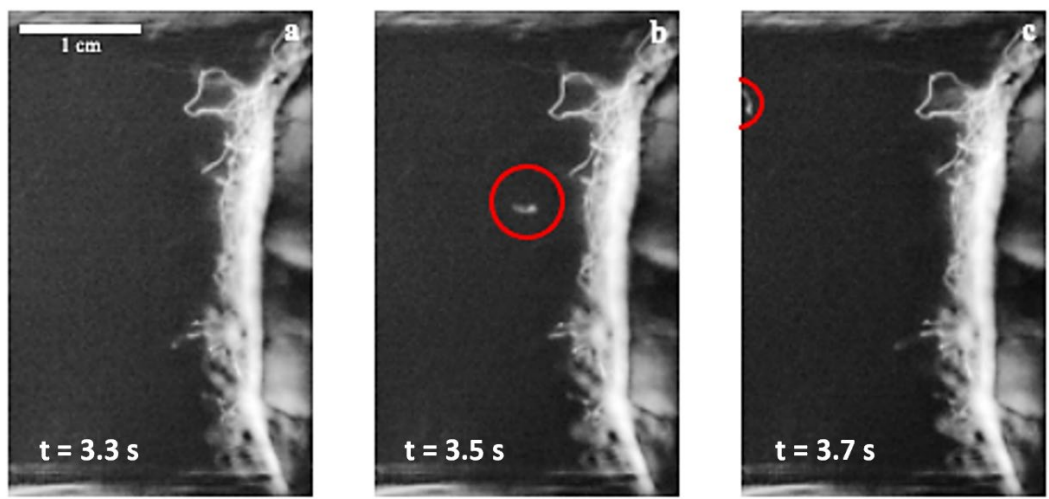

**Figure 2**. Example of a sequence illustrating one fiber released from a brand-new disposable face mask during an airflow rate of 45 L/min (red circles show the same fiber).

Masks are protective equipment against respiratory infections, especially for medical and healthcare workers, and play a crucial role in preventing and reducing the spread of the SARS-CoV-2 virus during the COVID-19 pandemic. During this global emergency, they protected the general population and supported local economies as many economic activities resumed their operations under the mask mandates. It is alarming, however, to see from our results how easy it is for fibers and particles to come off from the masks, bearing in mind that the materials used for surgical masks are all synthetic non-degradable plastic designed to last for a long time.

While surgical masks are designed for single-use and proper disposal, the unprecedented global demand and supply shortages during the COVID-19 pandemic have led people to reuse masks. This practice compromises the mask's intended protective effect (Feng et al., 2020) and increases the risk of microplastic release and inhalation. Although mask mandates for the general public have ceased in most countries since 2022, surgical masks continue to be indispensable protective tools for medical and healthcare professionals who are required to wear them for extended periods while on duty. Additionally, individuals have chosen to wear masks during infectious disease seasons and at indoor and public events as an added precautionary measure.

Considering these factors, it is crucial to prioritize the development of innovative surgical masks that minimize exposure to microplastics while addressing the environmental impact associated with their disposal. Such advancements should aim to enhance both the protective capabilities of the masks and their sustainability by using materials that are safe for users and friendly with the environment. Innovative surgical masks can better meet the needs of healthcare workers and the general population by striking a balance between efficacy, comfort, and environmental responsibility.

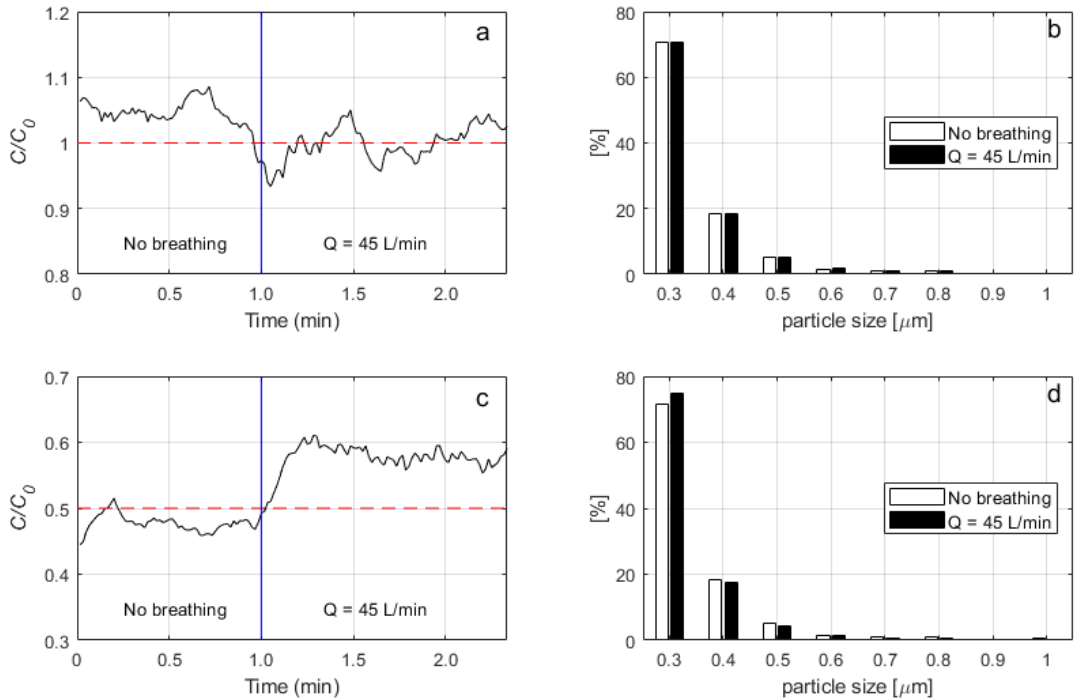

**Figure 3**. Particle detection using the OPS attached to the oscillatory airflow system. a) Ambient PNC (*C*) in the laboratory room over time measured with no mask and the system off (no breathing) and on at Q = 45 L/min. The average particle concentration during system operation with no mask ($C_0$) was used to normalize other concentrations. b) Particle-size distribution with no mask. c) Concentration of particles over time with the mask and the system on and off. d) Particle-size distribution with a mask.

4. Conclusions

This experimental investigation sheds light on the possibility of inhaling microplastics from disposable face masks, specifically focusing on surgical masks. Although fiber masks offer protective efficacy, they also pose an elevated risk of particle release and subsequent inhalation. While the exact implications are still not fully understood, the presence of microplastics in the human respiratory system is known to have a discernible impact on pulmonary function. This concern is particularly relevant for healthcare professionals who wear masks for prolonged periods and individuals who opt to use masks as an additional safeguard during infectious disease seasons or public gatherings. Addressing these concerns requires the development of innovative surgical masks. Such masks should effectively reduce exposure to microplastics while also considering the environmental consequences of plastic disposal. By incorporating novel concepts, like those proposed by, e.g. Yuk et al. (2022) and exploring new materials, as well as implementing sustainable practices such as the utilization of biodegradable or recyclable components, it is possible to manufacture masks that offer adequate protection while minimizing potential risks to human health and the environment.

Achieving an acceptable equilibrium between the protective capabilities of surgical masks and their impact on human health and the planet is imperative. Further research, regulatory measures, and advancements in mask design, usage guidelines, and disposal protocols are essential for ensuring the well-being and safety of individuals, particularly healthcare workers, while mitigating the adverse environmental effects of mask usage.

**Author Contributions Statement**

**Andres Prada:** Methodology, Investigation, Formal Analysis, Data Curation, Visualization, Writing- Original draft. **Avram Distler:** Investigation, Data Curation, Writing- Original draft. **Shyuan Cheng:** Methodology, Investigation, Writing- Reviewing and Editing. **John W Scott:** Conceptualization, Methodology, Supervision. **Leonardo Chamorro:** Conceptualization, Methodology, Writing- Original draft, Supervision. **Ganesh Subramanian:** Validation, Investigation, Data Curation, Visualization, Writing- Reviewing and Editing. **Vishal Verma:** Methodology, Resources, Writing- Reviewing and Editing, Supervision. **Andrew Turner:** Conceptualization, Writing - Review & Editing.

**References**

Akalin M, Usta I, Kocak D, Ozen M (2010) Investigation of the filtration properties of medical masks, Medical and Healthcare Textiles, Elsevier. pp. 93–97.

Atis S, Tutluoglu B, Levent E, Ozturk C, Tunaci A, Sahin K, Sarale A, Oktay I, Kanik A, Nemery B (2005) The respiratory effects of occupational polypropylene flock exposure. Eur Respir J. 25: 110–7. doi: 10.1183/09031936.04.00138403.

Basu S (2021) Computational characterization of inhaled droplet transport to the nasopharynx. Sci Rep 11: 6652. doi: 10.1038/s41598-021-85765-7.

Bayersdorfer J, Giboney S, Martin R, Moore A, Bartles R (2020) Novel manufacturing of simple masks in response to international shortages: Bacterial and particulate filtration efficiency testing. Am J Infect Contr 48: 1543−1545. doi: 10.1016/j.ajic.2020.07.019.

Chen Q, Gao J, Yu H, Su H, Yang Y, Cao Y, Zhang Q, Ren Y, Hollert H, Shi H, Chen C, Liu H (2022) An emerging role of microplastics in the etiology of lung ground glass nodules. Environ Sci Europe 34: 25. doi: 10.1186/s12302-022-00605-3.

Donaldson K, Brown D, Clouter A, Duffin R, MacNee W, Renwick L, Tran L, Stone V (2002) The pulmonary toxicology of ultrafine particles. J Aerosol Med 15(2): 213-20. doi: 10.1089/089426802320282338.

Feng S, Shen C, Xia N, Song W, Fan M, Cowling BJ (2020) Rational use of face masks in the COVID-19 pandemic. Lancet 8(5): 434-436. doi: 10.1016/S2213-2600(20)30134-X.

Huang S, Huang X, Bi R, Guo Q, Yu X, Zeng Q, Huang Z, Liu T, Wu H, Chen Y, Xu J, Wu Y, Guo P (2022) Detection and analysis of microplastics in human sputum. Environ Sci Tech 56: 2476−2486. doi: 10.1021/acs.est.1c03859.

He X, Reponen T, McKay RT, Grinshpun SA (2013) Effect of particle size on the performance of an N95 filtering facepiece respirator and a surgical mask at various breathing conditions. Aerosol Sci Technol. 47: 1180–1187. doi: 10.1080/02786826.2013.829209.

Jenner LC, Rotchell JM, Bennett RT, Cowen M, Tentzeris V, Sadofsky LR (2022) Detection of microplastics in human lung tissue using µFTIR spectroscopy. Sci Total Environ 831, 154907. doi: 10.1016/j.scitotenv.2022.154907.

Ju JTJ, Boisvert LN, Zuo YY (2021) Face masks against COVID-19: Standards, efficacy, testing and decontamination methods. Adv Colloid Interfac 292: 102435. doi: 10.1016/j.cis.2021.102435.

Kang PK, Shah DO (1997) Filtration of nanoparticles with dimethyl-dioctadecyl-ammonium bromide treated microporous polypropylene filters. Langmuir 13(6): 1820–1826. doi: 10.1021/la961010+.

Li L, Zhao X, Li Z, Song K (2021) COVID-19: Performance study of microplastic inhalation risk posed by wearing masks. J Hazard Mater 411: 124955. doi: 10.1016/j.jhazmat.2020.124955.

Matuschek C, Moll F, Fangerau H, Fischer JC, Zänker K, van Griensven M, Schneider M, Kindgen-Milles D, Knoefel WT, Lichtenberg A, Tamaskovics B, Djiepmo-Njanang FJ, Budach W, et al. (2020) The history and value of face masks. Eur J Med Res 25:23. doi: 10.1186/s40001-020-00423-4.

O'Dowd K, Nair KM, Forouzandeh P, Mathew S, Grant J, Moran R, et al. (2020) Face masks and respirators in the fight against the COVID-19 pandemic: a review of current materials. Adva Future Perspect Mater 13(15): 3363. doi: 10.3390/ma13153363.

Pauly JL, Stegmeier SJ, Allaart HA, Cheney RT, Zhang PJ, Mayer AG, Streck RJ (1998) Inhaled cellulosic and plastic fibers found in human lung tissue. Cancer Epiderm Biomar 7(5): 419–428.

Shokri A, Golbabaei F, A. Seddigh-Zadeh A, Baneshi MR, Asgarkashani N, Faghihi-Zarandi A (2015) Evaluation of physical characteristics and particulate filtration efficiency of surgical masks used in Iran's hospitals. Int J Occup Hyg 7(1): 10–16.

Spasiano D, Pirozzi F (2017) Treatments of asbestos containing wastes. J Environ Manag 204: 82-91. doi: 10.1016/j.jenvman.2017.08.038.

Strasser BJ, Schlich T (2020) A history of the medical mask and the rise of throwaway culture. Lancet 396(10243): 19-20. doi: 10.1016/S0140-6736(20)31207-1.

Tcharkhtchi A, Abbasnezhad N, Zarbini Seydani M, Zirak N, Farzaneh S, Shirinbayan M (2021) An overview of filtration efficiency through the masks: Mechanisms of the aerosol penetration. Bioact Mat. 6: 106–122. doi: 10.1016/j.bioactmat.2020.08.002.

Vianello A, Jensen RL, Liu L, Vollertsen J (2019) Simulating human exposure to indoor airborne microplastics using a Breathing Thermal Manikin. Sci Reports 9:8670. doi: 10.1038/s41598-019-45054-w.

Yang X, Yan Y, Xue C, Du X, Ye Q (2018) Association between increased small airway obstruction and asbestos exposure in patients with asbestosis. Clin Respir J 12(4): 1676-1684. doi: 10.1111/crj.12728.

Yao B, Wang Y, Ye X, Zhang F, Peng Y (2019) Impact of structural features on dynamic breathing resistance of healthcare face mask. Sci. Total Environ. 689: 743–753. doi: 10.1016/j.scitotenv.2019.06.463.

Yuk J, Chakraborty A, Cheng S, Chung C-I, Jorgensen A, Basu S, Chamorro LP, Jung S (2022) On the design of particle filters inspired by animal noses. J R Soc. Interface 19: 20210849. doi: 10.1098/rsif.2021.0849.

Zhang Q, Zhao Y, Du F, Cai H, Wang G, Shi H (2020) Microplastic fallout in different indoor environments. Environ Sci Technol (54) 6530−6539. doi: 10.1021/acs.est.0c00087.

Zhao M, Liao L, Xiao W, Yu X, Wang H, Wang Q, et al. (2020) Household materials selection for homemade cloth face coverings and their filtration efficiency enhancement with Triboelectric charging. Nano Lett 20: 5544–52. doi: 10.1021/acs.nanolett.0c02211.